\documentclass[aip,reprint]{revtex4-1}

\usepackage{graphicx}
\usepackage{amsmath}
\usepackage{amssymb}
\usepackage{caption}
\usepackage{subcaption}
\usepackage{color}
\usepackage{mathtools}
\usepackage{dcolumn}

\draft 

\begin{document}


\title{Uncertainty Quantification in Atomistic Simulations of Silicon using Interatomic Potentials} 



\author{I. R. Best}
\affiliation{School of Engineering, University of Warwick, Coventry, CV4 7AL, United Kingdom}

\author{T. J. Sullivan}
\affiliation{Warwick Mathematics Institute, University of Warwick, Coventry, CV4 7AL, United Kingdom}
\affiliation{School of Engineering, University of Warwick, Coventry, CV4 7AL, United Kingdom}
\author{J. R. Kermode}
\affiliation{School of Engineering, University of Warwick, Coventry, CV4 7AL, United Kingdom}


\date{\today}

\begin{abstract}
Atomistic simulations often rely on interatomic potentials to access greater time- and length- scales than those accessible to first principles methods such as density functional theory (DFT). However, since a parameterised potential typically cannot reproduce the true potential energy surface of a given system, we should expect a decrease in accuracy and increase in error in quantities of interest calculated from simulations. Quantifying the uncertainty on the outputs of atomistic simulations is  thus an important, necessary step so that there is confidence in results and available metrics to explore improvements in said simulations. Here, we address this research question by forming ensembles of Atomic Cluster Expansion (ACE) potentials, and using Conformal Prediction with DFT training data to provide meaningful, calibrated error bars on several quantities of interest for silicon: the bulk modulus, elastic constants, relaxed vacancy formation energy, and the vacancy migration barrier. We evaluate the effects on uncertainty bounds using a range of different potentials and training sets.
\end{abstract}


\maketitle 


\section{Introduction}

An interatomic potential (IP) is any function $V(\mathbf{x})$ which maps atomic positions $\mathbf{x}$ to the potential energy surface (PES) of a given material. They are constructed with the goal of efficiently computing energies and forces on atoms, but at the cost of reduced accuracy when compared to reference quantum-mechanical methods such as Density Functional Theory (DFT). This loss of accuracy is regrettable but necessary in the computation of expensive Quantities of Interest (QoIs), which would otherwise be inaccessible due to the unfeasible scaling of these ab-initio techniques with system size or simulation time. Historically, IPs with fixed functional forms were tuned on a per-species and/or per-QoI basis \cite{StillingerWeber1985,DawBaskesEAM} with the specific form of the potential usually inspired by physical or chemical intuition. While these IPs are usually successful on the quantities and chemical systems on which they are trained, they typically have limited transferability to new species or quantities since the parameters and form of the model are highly specific, and targeting another QoI or material requires either re-tuning the model parameters, or building a new model from scratch.

The advent of Machine Learning Interatomic Potentials (MLIPs), trained on DFT or other high-accuracy data, explores the construction of systematically improvable approximations to the PES of a given material. These potentials \cite{Bartok2010,Thompson2015,Shapeev2016,Drautz2019,NEURIPS2022_4a36c3c5} promise accuracy and transferability \cite{Deringer2019} between different QoIs for an appropriately chosen and trained potential. However, it remains a difficult task to identify where a MLIP is extrapolating outside of observed training data, since the PES landscape is complex for even simple materials. Accurate estimation of the PES and transferability between QoIs is only likely in regions where a model is close to, or interpolating between, training data.

MLIPs are frequently leveraged within atomistic simulations to take advantage of the aforementioned accuracy and transferability, coupled with the availability of ab-initio materials science databases \cite{Becker2013,Draxl2019} on which to train. A high-accuracy input dataset from a more fundamental level of theory is used to train a potential, and this data is typically split into training and test sets - the performance of an MLIP is often evaluated using metrics like train and test error. These measures can be useful in diagnosing features of a trained MLIP, an example being to characterise the level of over-fitting to training data; nevertheless, these error metrics should not be treated as uncertainty measures. Even a machine learned potential which achieves low training and test errors can perform poorly when extrapolating outside of previously observed configurations. These error metrics are also calculated on quantities already in the training set, and not on the QoIs which the model will eventually be used to predict: there is no clear connection or conversion between these metrics and an uncertainty in a given QoI. Whilst we might hope that a model with low train and test errors will perform well on a range of different QoIs, this is not a guarantee since our train and test datasets do not cover all possible atomic configurations.

Uncertainty quantification (UQ) in the context of atomistic simulation must therefore account for several different sources of uncertainty - for example, noise on underlying training data, parametric uncertainty, and model form error - and combine these different sources in a statistically meaningful way. Previous work inspired by Bayesian frameworks has constructed ensembles of fixed form IPs by sampling coefficients from a cost landscape \cite{Frederiksen2004,Longbottom_2019}, weighting different possible choices of coefficients with a fictitious temperature to assess the parametric uncertainty in fixed form and machine-learned IPs, and analysing the `sloppiness' in model parameters through different statistical lenses \cite{Kurniawan2022}. For MLIPs, several different frameworks have been used to estimate uncertainties. As a non-exhaustive list of examples: Gaussian Process Regression based UQ via cross validation \cite{Bartok2022} and subsampling \cite{doi:10.1021/acs.jctc.8b00959}, uncertainties using ACE potentials to drive active learning of datasets \cite{VanderOord2022,Lysogorskiy2023}, ensembles with neural networks \cite{Carrete2023} (NNs), dropout NNs with a Bayesian interpretation \cite{Wen2020}, and conformal prediction with NNs on quantities within the training set as a comparison to other methods \cite{Hu_2022} and to calibrate uncertainties in building potentials with MD \cite{zaverkin2023uncertainty}. 

In an attempt to build upon these previous approaches to UQ in atomistic simulation, in this paper we aim to provide robust uncertainty estimates on a range of QoIs calculated with a variety of different Atomic Cluster Expansion (ACE) \cite{Drautz2019} potentials for pure silicon, trained on a reference DFT dataset \cite{Bartok2018}. The choice of potential family to ACE is made to take advantage of the linear structure of the IP, which is advantageous both for evaluation speed \cite{Lysogorskiy2021} and to enable efficient statistical sampling - it is conceptually easier to place priors on coefficients of a linear model, rather than considering classical IPs or some differently architectured MLIP, where there may exist some hierarchy of importance in coefficients. 

We setup our linear ACE basis inside a Bayesian inverse problem, which after accounting for parametric uncertainty and noise within our training data, provides a posterior distribution of coefficients from which we can form an ensemble of ACE potentials - here the speed of energy and force evaluations becomes important, since if we desire statistical significance we require many evaluations of a given QoI, which can translate to an extremely large amount of atomic observation calculations. We push samples from the ensemble through the calculation of a QoI, and calibrate the observed uncertainties using conformal prediction \cite{Angelopoulos2021}. This workflow can be readily translated to other classes of MLIPs, different QoIs and different sources of training data.

\section{Methodology} 

\subsection{ACE Potentials}
\label{sec:ace potentials}

The Atomic Cluster Expansion exploits a body order expansion to construct potentials of a desired correlation order $\nu$ ($=\text{body order} - 1$) which are naturally invariant to translations, rotations and permutations of same species atoms. An outline of the ACE construction is provided below; for more details, see e.g. \cite{Drautz2019,DUSSON2022110946,witt2023acepotentials}.

The core of an ACE model is the projection of the atomic density $\rho$
\begin{equation}
	\rho(\mathbf{r}) = \sum_{j \neq i}\delta(\mathbf{r} - \mathbf{r}_{ji})
	\label{atomic density}
\end{equation}
onto the one-particle basis $\varphi_{v}$
\begin{equation}
    \varphi_{v}(\mathbf{r}_{ij}) = R_{nl}(r_{ij})Y_{l}^{m}(\hat{r})
    \label{ACE one particle basis}
\end{equation}
where $v = (nlm)$ is a multi-index of the principal, orbital and magnetic quantum numbers, $\mathbf{r}_{ij} = r_{ij}\hat{r}$ is the distance vector of magnitude $r_{ij}$ and direction $\hat{r}$, and $R_{nl}(r_{ij})$ and $Y_{l}^{m}(\hat{r})$ are the radial functions and the spherical harmonics respectively. These $\varphi$ are evaluated for the $N(i)$ neighbours of a given atom $i$ within some cutoff region. To then construct a  description/basis of the atomic environment which is invariant to permutations, we perform a summation of all the $\varphi$ within this cutoff  
\begin{equation}
    A_{iv} = \sum_{j \in N(i)} \varphi_{v}(\mathbf{r}_{ij}) .
    \label{ACE atomic base}
\end{equation}
and then form the product A-basis
\begin{equation}
    \mathbf{A}_{i,\mathbf{v}} = \prod_{\gamma}^{\nu} A_{iv_\gamma}
    \label{A basis}
\end{equation}
with $\mathbf{v} = (v_1,v_2,...,v_{\nu})$ and $\nu$ the correlation order, or equivalently the $(\nu+1)$ body order of the expansion. 

Performing a body-order expansion of the site energy $V_i$ (which can also be done equivalently with other quantities) with this $A$ basis leads to
\begin{equation}
    \begin{split}
        V_i &= V_i^{(0)} + \sum_{v} \tilde{c}_v^{(1)}A_{iv}
        + \sum_{v_1 v_2}^{v_1 \geq v_2} \tilde{c}_{v_1 v_2}^{(2)}A_{iv_1}A_{iv_2} \\
        &+ \sum_{v_1 v_2 v_3}^{v_1 \geq v_2 \geq v_3} \tilde{c}_{v_1 v_2 v_3}^{(3)}A_{iv_1}A_{iv_2}A_{iv_3} + ... \hspace{0.2cm} .
    \end{split}
    \label{body order expansion A basis}
\end{equation}

In order for this basis to be rotationally invariant, it is necessary to average over the three-dimensional rotations and remove those basis functions which are zero or linearly dependent, i.e. those encoding the same information, which leads to the $B$ basis
\begin{equation}
    \mathbf{B}_{i,\mathbf{v}} = \sum_{\mathbf{v}'} \mathcal{C}_{\mathbf{v}\mathbf{v}'} \mathbf{A}_{i,\mathbf{v}'}
    \label{ACE B basis}
\end{equation}
with $\mathcal{C}_{\mathbf{v}\mathbf{v}'}$ the sparse Clebsch-Gordon coefficient matrix.

An energy evaluation using ACE is then given by
\begin{equation}
    V(\mathbf{c},\mathbf{R}) = \sum_i c_{i,\mathbf{v}} \mathbf{B}_{i,\mathbf{v}}(\mathbf{R}) = \mathbf{c} \cdot \mathbf{B}(\mathbf{R})
    \label{ACE energy evaluation}
\end{equation}
where $\mathbf{R} = (\mathbf{r}_1, \mathbf{r}_2, ..., \mathbf{r}_n)$ are the positions of the atoms (and, in practice, the unit cell of the structure $\mathbf{R}$ and numbering of the species of atom for each $\mathbf{r}_i$, omitted here for clarity). The linear nature of the potential is now obvious, and the parameters $\mathbf{c}$ are the ones that must be determined by fitting to data. 

The important properties of the potential for this paper are that we have a flexible, accurate \cite{Drautz2019,Qamar2023} and efficient \cite{DUSSON2022110946,Lysogorskiy2021} MLIP, and once we specify $\nu$ and the maximum total polynomial degree $p_{\text{max}}$ (controlling the number of coefficients in our potential, by picking out all basis functions such that $p_{\text{max}} \geq \sum_{i=1}^{\nu} n_i + l_i$ , see\cite{DUSSON2022110946,witt2023acepotentials}), as well as relevant cutoffs and weights on different observation types, then atomic properties - energies, forces and virial stresses - are linear combinations of the potential coefficients basis functions (and relevant derivatives). These atomic properties can then be fed into any more complex QoI evaluation and propagated through a UQ procedure. 

We utilise the ACEpotentials.jl Julia package \cite{ACEpotentials.jl} for construction of all potentials and evaluations of atomic properties. The potentials in this paper use a cubic pair potential term for both the short range repulsive behaviour and long distance behaviour up to final (smooth) cutoff of $3.0r_0$ (where $r_0 \approx 2.35$~\AA{} is an estimate of the nearest neighbour distance in silicon), and a many body ACE part (of order $p_\mathrm{max}$) for the intermediate ($0.8r_0 - 2.0r_0$) regime. We also set the correlation order $\nu = 3$, i.e. we include up to 4-body terms in the expansion, and set the weights on our differently labelled training data as per the supplementary information of Ref.~\onlinecite{Lysogorskiy2021}. This set-up is not intended to be well optimised for the problems considered below; this methodology can be applied to any linear potential constructed in a similar manner, and is not specific to these design choices.

\subsection{Bayesian Inverse Problems}
\label{bayesian inverse problems}

Begin with some generic data $(\mathbf{x},\mathbf{y})$, both of length $N$. We can view this data as atomic positions of different structures $\mathbf{x} = (\mathbf{R}_1, \mathbf{R}_2, ..., \mathbf{R}_N)$, and the corresponding DFT energies, forces and virials $\mathbf{y} = (\mathbf{y}_1, \mathbf{y}_2, ..., \mathbf{y}_N)$ where $\mathbf{y}_i$ consists of a single energy, six virial stresses and $3n$ forces with $n$ the number of atoms in $\mathbf{R}_i$. 

Utilising the linear nature of the ACE potentials described previously, we proceed from Equation \ref{ACE energy evaluation} and set up our data problem in the following way:
\begin{equation}
    \mathbf{y} = V(\mathbf{c},\mathbf{x}) + \boldsymbol{\epsilon} \hspace{2.5pt},\hspace{2.5pt} \boldsymbol{\epsilon} \sim \mathcal{N}(\textbf{0},\beta^{-1}\mathbb{I}) ,
\end{equation}
where we assume that the potential $V$ is capable of reproducing our targets $\mathbf{y}$ (the `ground truth') which are contaminated with Gaussian, zero mean noise $\boldsymbol{\epsilon}$, with homoscedastic precision $\beta$, representing the noise we estimate on our data.

We wish to find a distribution describing the parameters (or coefficients) $\mathbf{c}$ of the model $V$ through a basic Bayesian inverse problem. Utilising Bayes' rule,
\begin{equation}
    \mathbb{P}(\mathbf{c}|\mathbf{y},\alpha,\beta) = \frac{\mathbb{P}(\mathbf{y}|\mathbf{x},\mathbf{c},\beta) \mathbb{P}(\mathbf{c}|\alpha)}{\mathbb{P}(\mathbf{y})},
    \label{Bayes Rule}
\end{equation}
our posterior distribution of weights given data, $\mathbb{P}(\mathbf{c}|\mathbf{y},\alpha,\beta)$, is the product of the likelihood of observing the data $\mathbb{P}(\mathbf{y}|\mathbf{x},\mathbf{c},\beta)$ and the prior for our parameters $\mathbb{P}(\mathbf{c}|\alpha)$ given some assumed precisions on parameters $\alpha$ and data $\beta$, normalised by the evidence $\mathbb{P}(\mathbf{y})$. Here we assume a single shared precision hyperparameter $\alpha$ for all the coefficients in the model.

To proceed from Equation \ref{Bayes Rule}, we must either sample from our posterior $\mathbb{P}(\mathbf{c}|\mathbf{y},\alpha,\beta)$ using e.g. MCMC, or have an analytical form for the posterior distribution. Forcing analyticity, while simpler mathematically, is not the most general approach, however this is not expected to be a problem for most atomistic systems; as the amount of training data is typically quite large, and therefore the posterior will be less sensitive to the chosen prior. 

Also of note here is the divergence between the construction of the relatively simple potentials within this paper and the current state of the art for ACE potentials (as well as the discussion on the construction choices in Section \ref{sec:ace potentials}). Construction of state of the art ACE potentials \cite{witt2023acepotentials} utilise \textit{smoothness priors} to impose smoothness on the potential fit - one possible choice to perform this regularisation in the fitting procedure is to specify the prior distribution $\mathbb{P}(\mathbf{c}|\alpha)$, and can be compared to performing Bayesian ridge regression / regularised least squares - however this is only one possibility: for more details, refer to the recent overview of ACEpotentials.jl.\cite{witt2023acepotentials}

We can write the likelihood of our model as a multivariate normal distribution
\begin{equation}
    \mathbb{P}(\mathbf{y}|\mathbf{x},\mathbf{c},\beta) = \mathcal{N}(\mathbf{y}|V(\mathbf{c},\mathbf{x}),\beta^{-1}\mathbb{I}) ,
    \label{mv likelihood}
\end{equation}
and we choose a conjugate prior for the parameters 
\begin{equation}
    \mathbb{P}(\mathbf{c}|\alpha) = \mathcal{N}(\mathbf{0},\alpha^{-1}\mathbb{I}) ,
    \label{conjugate prior}
\end{equation}
such that we can construct our posterior analytically:
\begin{equation}
    \mathbb{P}(\mathbf{c}|\mathbf{y},\alpha,\beta) = \mathcal{N}(\mathbf{c}|\boldsymbol{\mu},\mathbf{S}) ,
    \label{weight posterior}
\end{equation}
where
\begin{equation}
    \mathbf{S} = (\alpha \mathbb{I} + \beta \Phi^T \Phi)^{-1} , \hspace{0.5cm} \boldsymbol{\mu} = \beta \mathbf{S} \Phi^T \mathbf{y}\hspace{0.2cm}.
\end{equation}

Here, the mean vector and covariance matrix are parameterised in terms of our target data $\mathbf{y}$, precision hyperparameters $\alpha$ and $\beta$, and a design matrix
\begin{equation}
    \Phi = 
    \begin{pmatrix}
   B_1 \left(\mathbf{R}_1\right) & B_2 \left(\mathbf{R}_1\right) & ... & B_M \left(\mathbf{R}_1\right) \\
   B_1 \left(\mathbf{R}_2\right) & B_2 \left(\mathbf{R}_2\right) & ... & B_M \left(\mathbf{R}_2\right) \\
   \vdots & \vdots & \ddots & \vdots\\
   B_1 \left(\mathbf{R}_N\right) & B_2 \left(\mathbf{R}_N\right) & ... & B_M \left(\mathbf{R}_N\right)
    \end{pmatrix},
    \label{design matrix}
\end{equation}
the entries of which evaluate the j\textsuperscript{th} basis function on the i\textsuperscript{th} data point.

\subsection{Hyperparameters and optimisations}
\subsubsection{Hyperparameters and the Evidence function}

We have introduced two hyperparameters into the inverse problem: a single noise precision $\beta$ which estimates the noise inherent in our training data $\mathbf{y}$, and a single weight precision $\alpha$ which estimates our confidence in the possible model parameter values. 

A single, homoscedastic noise precision value $\beta$ assumes that our training data are independent draws from the same underlying distribution. While the training data in our case $(\mathbf{x},\mathbf{y})$ are atomic configurations and DFT observations from the same database, they are not drawn from the same distribution - $\mathbf{y}$ contains related but markedly different quantities (total energies, forces on atoms, and virial stress components). The single $\beta$ assumption is an extreme simplification - introducing heteroscedasticity between different types of quantities contained within $\mathbf{y}$, or further having different $\beta$ for different atomic configurations $\mathbf{x}$, would be a far more accurate approach at the expense of a higher dimensional hyperparameter space. 

Similarly, the single weight precision value $\alpha$ assumes similar confidence in all coefficients $\mathbf{c}$ in the chosen model. This is unlikely to be true, especially in the construction of ACE potentials in this paper as we have pair potential and many-body terms included in the linear basis. An extension which would follow naturally through Equations \ref{Bayes Rule}-\ref{weight posterior} is the promotion of $\alpha$ from scalar to vector-valued, which would allow more flexibility in the weight posterior (\ref{weight posterior}) and further analysis (for example, in the form of Automatic Relevance Detection, ARD), again at the cost of increased complexity of the inverse problem and subsequent hyperparameter optimisations and sampling.

In a fully Bayesian treatment, these hyperparameters should be sampled over by setting suitable hyperpriors, and then sampling the entire Bayesian posterior. For simplicity, we instead optimise $\alpha$ and $\beta$ such that they maximise the (log-) evidence \cite{bishopmachinelearning} or log-marginal likelihood: the likelihood of observing our data $\mathbf{y}$ given a model defined by hyperparameters $\alpha$ and $\beta$. This is given by
\begin{equation}
\begin{split}
    \ln \mathbb{P}(\mathbf{y}|\alpha, \beta) &= \frac{M}{2}\ln \alpha + \frac{N}{2} \ln \beta - E(\mathbf{c}) \\
    &\hspace{0.5cm}- \frac{1}{2}\ln |\mathbf{S}^{-1}| - \frac{N}{2}\ln 2\pi ,
\end{split}
\label{log evidence}
\end{equation}

where 
\begin{equation}
    E(\mathbf{c}) = \frac{\beta}{2}|| \mathbf{y} - \Phi \mathbf{c} ||^2 + \frac{\alpha}{2} \mathbf{c}^T \mathbf{c}  .
    \label{log evidence cost?}
\end{equation}

Note the trade-off in Equation \ref{log evidence cost?} between reproducing the training data and the penalisation for complicated models with many parameters. For simple models with small coefficient vector $\mathbf{c}$ (length $M$), agreement between targets $\mathbf{y}$ and prediction $\Phi \mathbf{c}$ will likely be poor, and the evidence will be low. For larger $M$ this agreement will be better but the model evidence is penalised for the added complexity, due to the increase in size of $\mathbf{c}$ and the inner product $\mathbf{c}^T \mathbf{c}$. This is one method of analysing the level of over-fitting when comparing different models: the value of $\ln \mathbb{P}(\mathbf{y}|\alpha, \beta)$ for increasing size of some model $V$ should have a maximum value, past which the increasing performance on the training set can no longer compensate for the increasing complexity of the model.

\subsubsection{Optimisation with Particle Swarm}

Now we must choose an algorithm to maximise Equation \ref{log evidence} with respect to $\alpha$ and $\beta$. There exist iterative update schemes \cite{MacKay1992,bishopmachinelearning} to perform this, but they scale and perform poorly as $\Phi$ becomes larger - the hyperparameter space becomes more complex, and individual evidence evaluations become slower as the corresponding matrices in Equation \ref{log evidence} grow in size. A gradient based approach like an L-BFGS performs more efficiently, but is sensitive to initial point chosen during optimisation and can fail if it reaches a very unlikely position in the hyperparameter landscape. This is made worse since we are generating many different potentials throughout this paper - either in terms of varying types and amounts of training data, and/or vastly different numbers of coefficients. It becomes difficult to make a meaningful ansatz for sensible initial precision hyperparameter values which can span several orders of magnitude for different models. 

To circumvent these problems, we utilise a particle swarm optimisation (PSO) algorithm. This global optimisation scheme initialises a population of particles with random velocities across the hyperparameter space $(\text{log} \hspace{0.06cm} \alpha,\text{log}\hspace{0.06cm} \beta)$, where we operate in a log- space since precisions must be positively valued and can vary over many orders of magnitude. At each step, each particle is accelerated towards the position of the global best value for the (log-) evidence, and the position of the best value the individual has observed. 

By tuning these relative accelerations - as well as other parameters of the optimisation like the number and inertia of the particles in the swarm, and the number of iterations - the algorithm can balance local and global exploration of the hyperparameter landscape and relatively quickly converge to a sensible result for $\alpha$ and $\beta$. Individual iteration steps for each particle do not require gradient evaluations, however the calculation of the (log-) evidence (\ref{log evidence}) must occur for each member of the swarm at each iteration step. 

We do not have a guarantee of convergence to a local or global optimum using PSO, but in application the swarm regularly converges for the precision hyperparameters $\alpha$ and $\beta$ for a range of different potentials and training data for a swarm of 50 particles allowed to proceed for 200 iterations, with equal weightings for local and global exploration and an overall inertia weight of 0.9 to prevent divergence of swarm members. On the occasions for which PSO fails/diverges, the values observed for $\alpha$, $\beta$ and the evidence are easily identified as unphysical, and PSO can usually be re-run without changing the algorithm settings. 

\subsection{Conformal Predictions}

If we perform our parameter estimation as a Bayesian inverse problem as in Section \ref{bayesian inverse problems}, we operate under the assumptions that the noise in our data and the prior and posterior distributions for our coefficients are normally distributed, and that the model we are using is the correct one, i.e., there is no model form error. While the distributional assumptions might be allowable, there will almost always be some amount of model form error for an IP attempting to match DFT training and test data (e.g. due to limited body order and cutoff distances), and there are other sources of uncertainty which we will not explicitly account for in this workflow (for instance, algorithmic uncertainty in calculating specific QoIs).

Furthermore, if we wish to transform our weight posterior $\mathbb{P}(\mathbf{c}|\mathbf{y},\alpha,\beta)$ into a distribution of output quantities $\mathbf{y}_{\text{new}}$ of the same type as $\mathbf{y}$ (so in our case, containing energies, forces and virial stresses, on the new atomic configurations $\mathbf{x}_{\text{new}}$), and from these calculate a given QoI, we can do so using a predictive posterior distribution given by \cite{bishopmachinelearning}
\begin{equation}
    \mathbb{P}(\mathbf{y}_{\text{new}}|\mathbf{y},\alpha,\beta) = \int 
    \mathbb{P}(\mathbf{y}_{\text{new}}|\mathbf{x},\mathbf{c},\beta) \mathbb{P}(\mathbf{c}|\mathbf{y},\alpha,\beta)\;\mathrm{d}\mathbf{c},
\label{predictive posterior integral}
\end{equation}
which for our normal likelihood in Equation \ref{mv likelihood} and weight posterior in Equation \ref{weight posterior} can be evaluated analytically:
\begin{equation}
    \mathbb{P}(\mathbf{y}_{\text{new}}|\mathbf{y},\alpha,\beta) = \mathcal{N}(\boldsymbol{\mu}^T \mathbf{B}(\mathbf{x}_{\text{new}}), \Sigma)
\end{equation}
where
\begin{equation}
    \Sigma = \frac{1}{\beta} + \mathbf{B}(\mathbf{x}_{\text{new}})^T \mathbf{S} \mathbf{B}(\mathbf{x}_{\text{new}}) .
\end{equation}

This predictive posterior would in principle give the distributions in energies, forces and virial stresses which can be fed through a QoI calculation. However for atomistic simulations it is in general quite difficult to form, since we require foreknowledge of the atomic configurations $\mathbf{x}_{\text{new}}$ - for QoIs involving relaxations or dynamics under a given potential, this is not possible.


Instead, we are left with sampling from our weight posterior from Equation \ref{weight posterior} and pushing the corresponding potentials through a QoI simulation. This results in tight output QoI distributions, with very small predicted uncertainties, since we are not performing the integral in Equation \ref{predictive posterior integral}. The uncertainties we could calculate by doing this here are not the ones from a true Bayesian inverse problem, since we only propagate potential parameter vectors $\mathbf{c}$ from the weight posterior forward through the model. Nevertheless, the small uncertainties we predict from the forward propagation of ensemble members can still be useful, since the spread of predictions still has some of the correct qualitative behaviour we would like our uncertainty to capture more rigourously. 

We therefore aim to calibrate these uncertainties by performing conformal prediction,\cite{vovk2005algorithmic, Angelopoulos2021} which seeks to construct a distribution free, frequentist prediction set $\mathcal{C}$ such that the probability the true result lies within $\mathcal{C}$ is close to a set desired coverage $1-\zeta$, i.e.
 \begin{equation}
1 - \zeta \leq \mathbb{P}(\mathbf{y}_{\text{new}} \in \mathcal{C}(\mathbf{x}_{\text{new}})) \leq 1 - \zeta + \frac{1}{n+1} ,
\label{conformal prediction set}
\end{equation}
with $n$ the length of the calibration set and $(\mathbf{x}_{\text{new}},\mathbf{y}_{\text{new}})$ are points in a new set to be predicted from our model.  

The requirements for such a conformal procedure are a model $V(\mathbf{c},\mathbf{x})$ which outputs some mean prediction $\mu_i$ and some heuristic uncertainty measure $\sigma_i$ for an input $i$, as well as data $(\mathbf{x},\mathbf{y})$ split into training, calibration and test subsets. The only change compared to the procedure above is to set aside $n$ data points for the calibration of our error bars. To actually perform the calibration, we follow the procedure outlined in \cite{Angelopoulos2021}: 

\begin{enumerate}
    \item Define a score function $s(\mu_i,\sigma_i,y_{\text{cal},i})$ where larger scores encode worse agreement between the model prediction and the `true' value in the calibration set, in our case
\begin{equation}
s_i = \frac{\text{abs}(\mu_i - y_{\text{cal},i})}{\sigma_i} ,
\label{score function}
\end{equation}
and compute the scores on the calibration set
\begin{equation}
    \mathbf{s} = \{ s_i \}_{i=1}^n .
\end{equation}

    \item Compute $\hat{q}$, the multiplicative factor by which we will scale our uncertainties, from the list of scores $\mathbf{s}$
\begin{equation}
\hat{q} = \text{quantile}(q_{val},\mathbf{s}) \hspace{0.2cm}, \hspace{0.2cm} q_{val} = \Bigg\lceil {\frac{(n+1)(1-\zeta)}{n}} \Bigg\rceil .
\label{q hat and q val}
\end{equation}
    \item Form prediction sets on values in the new/prediction set:
\begin{equation}
\mathcal{C} = \left[ \mu_{\text{new}} - \hat{q}\sigma_{\text{new}}, \mu_{\text{new}} + \hat{q}\sigma_{\text{new}} \right] .
\label{prediction set C}
\end{equation}
\end{enumerate}
Note that when reporting values with uncertainties within this paper, we will often report them in the form $\mu_{\text{new}} \pm \hat{q}\sigma_{\text{new}}$ as a short form for a bound like Equation \ref{prediction set C}.

The form of Equation \ref{conformal prediction set} is a strong condition for a prediction set $\mathcal{C}$, however (as discussed elsewhere \cite{Angelopoulos2021}) much of the usefulness of $\mathcal{C}$ is tied to the chosen score function $s(\mu_i,\sigma_i,y_{\text{cal},i})$ and the make-up and size $n$ of the calibration set: a poor choice of score function, or a small calibration set, would result in a prediction set which in principle agrees with Equation \ref{conformal prediction set}, but in practice would not be a useful measure of the true uncertainty of the quantities in the prediction set of the trained model $V$. In this paper we will use the score function in Equation \ref{score function} with the heuristic uncertainty $\sigma$ being the standard deviation of the ensemble QoI prediction; refining this choice is an area of interesting further research into the choice of uncertainty measure $\sigma$ (e.g. standard deviation, interquartile range) and form of the score function, and the effects on the subsequent UQ.

We can verify that the conformal approach we employ has the correct coverage $C$ on these atomic properties by shuffling the calibration and prediction/test sets for a number of trials $R$, and plotting the distribution of the empirical coverage. As $R$ increases, we expect the mean empirical coverage to approach $1-\zeta$, and the distribution of empirical coverages to approach the analytic Beta-Binomial distribution \cite{Angelopoulos2021}
\begin{equation}
    C_{\text{analytic}} \sim \frac{1}{n_{\text{test}}} \text{BetaBin}(n_{\text{test}},n+1-l,l) ,
    \label{empirical coverage c_analytic}
\end{equation}
where $n_{\text{test}}$ is the length of the test set and 
\begin{equation}
    l = \lfloor (n+1)\zeta \rfloor .
\end{equation}

\begin{figure*}[!htbp]
	\centering
	\begin{subfigure}[b]{0.49\textwidth}
		\centering
		\includegraphics[width=\textwidth]{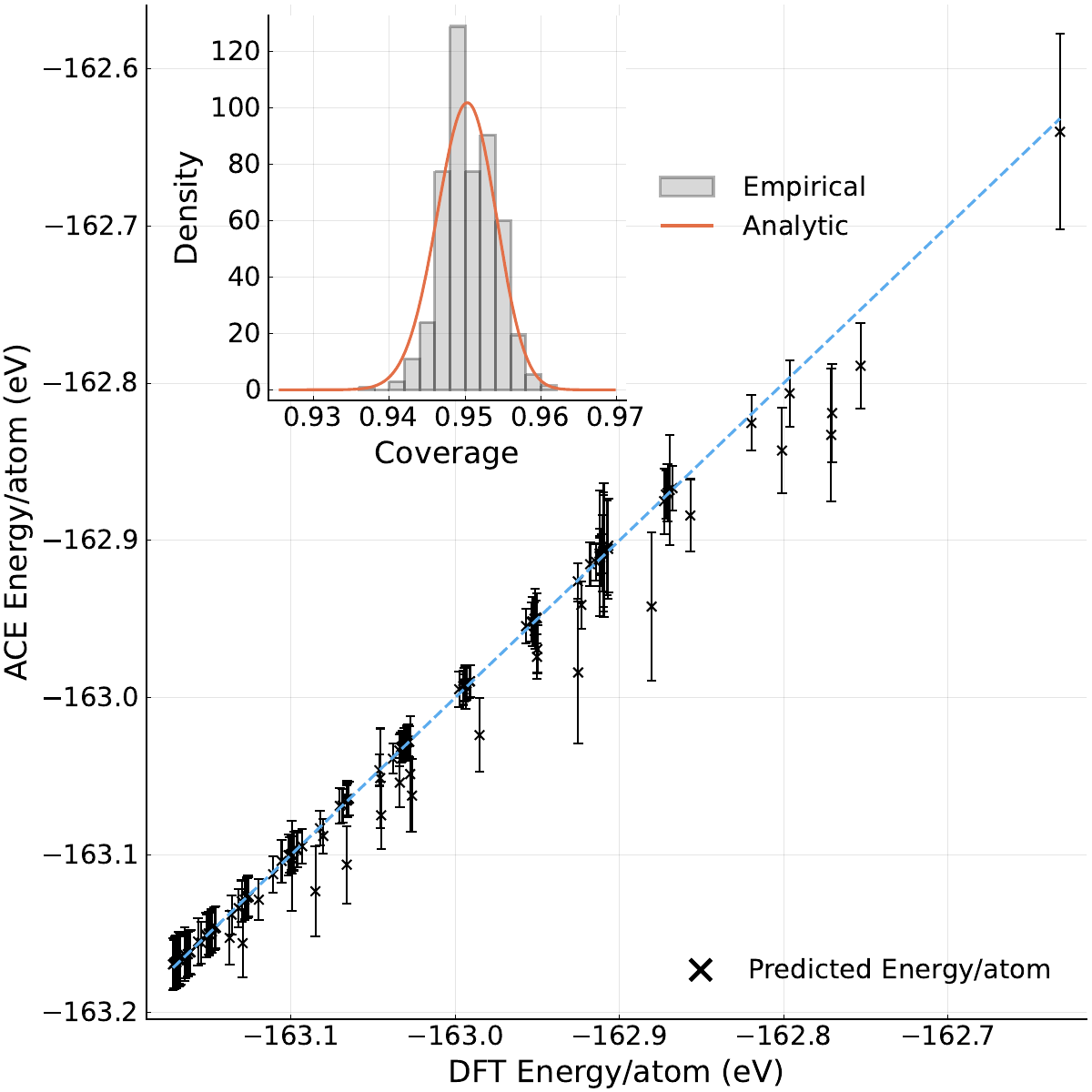}
	\end{subfigure}
	\hfill
	\begin{subfigure}[b]{0.49\textwidth}
		\centering
		\includegraphics[width=\textwidth]{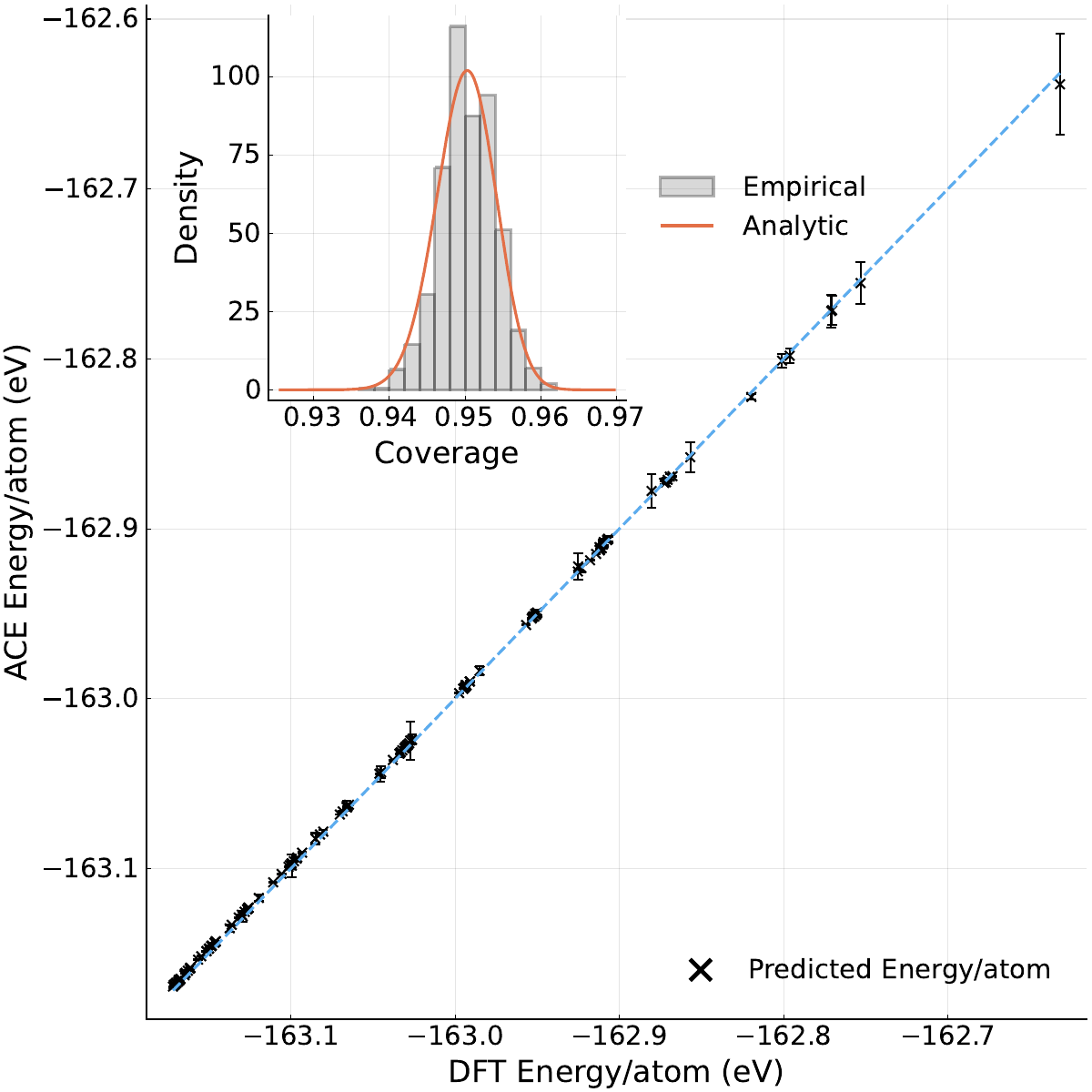}
	\end{subfigure}
	\caption{The energy per atom with conformal uncertainty bounds evaluated on the calibration and test sets for two differently sized potentials of 9 and 81 parameters, both trained on 300 bulk diamond configurations of silicon. The inset histogram compares the empirical observed conformal coverage for 1000 shuffles of the calibration and test sets with the analytic coverage distribution.}
	\label{true vs predicted E plus coverage dist}
\end{figure*}

For two simple ACE models of 9 and 81 parameters, both trained on 300 bulk configurations, calibrated on 39 bulk configurations and a test set of 98 bulk configurations, we plot the distribution of empirical coverages over 1000 trials in the insets of Figure \ref{true vs predicted E plus coverage dist}, and can follow this through by plotting uncertainties on these atomistic quantities, e.g. for the energy per atom of configurations in the calibration and test sets in Figure \ref{true vs predicted E plus coverage dist} - note that our coverage target includes energies, forces and virials, and does not guarantee coverage on solely e.g. energies (the coverage of energies within Figure \ref{true vs predicted E plus coverage dist} are 88\% and 83\% respectively on the 137 total structures). We can conclude that our conformal procedure is working as intended for the atomistic quantities in the training and calibration sets - we can place error bounds on energies, forces and virial stresses of the same form as Equation \ref{prediction set C}, with a coverage which we specify beforehand. 

It is important to make clear the practical differences between an ideal application of conformal prediction and the method we will develop here. A pure conformal approach calibrates the uncertainties of data \textit{on which the model is trained and calibrated}, which in our case are total energies, forces on atoms, and virial stresses. 

We extend the method above by \textit{assuming} we can also scale the uncertainty in a given QoI prediction from a sampled posterior by the same $\hat{q}$ value and obtain a sensible estimate of uncertainty - scaling the QoI spread of a model formed as in Equation \ref{weight posterior} (a `base' Bayesian model which accounts for parametric uncertainty and noise predicted on the data) by the performance of said model on the output of atomistic properties on which it was trained, leading to a frequentist prediction set for a QoI. We are also relaxing / ignoring the assumption that the points on which we make predictions $(\mathbf{x}_{\text{new}},\mathbf{y}_{\text{new}})$ are drawn from the same distribution as those in the training and calibration sets: this is already difficult in atomistic datasets in general due to the different types of observation present (e.g. energies, forces, virials, whilst related will be drawn from different underlying distributions) and different types of configurations (e.g. bulk, monovacancy, interstitials). In the case of differently labelled data, we also have the problem that a random split is not guaranteed to place members of a given label in all the subsets - train, calibrate and test - of the original data. In this paper we assume that members from each label are present in all the subsets, but there can be edge cases where there would be unexpected behaviour either in the trained potential or in the calibration of uncertainties, especially for a label with fewer members. In situations of data scarcity, whether from expense of simulation or otherwise, it may become necessary to explicitly control the split location of some data points.

It is the hope that this two stage UQ procedure leads to robust and interpretable error bounds on QoIs, and that this procedure can be adapted across a wide range of applications e.g. other forms of MLIPs, different formulations of the inverse problem (`base' models), other measures of heuristic uncertainty or score functions in the conformal procedure, and varied QoIs for different atomic systems. In application in this paper, following the prescription outlined in Equations \ref{score function}-\ref{prediction set C} with the same score function $s$, we set our desired coverage $1-\zeta = 0.95$ and split the available data such that approximately the train/calibrate/test split is $72:8:20$; as well as using an ensemble size of 100 members to both push through simulation and to calibrate our error bars. These choices are to a certain extent arbitrary: for example, the 95\% coverage choice is due to a traditional assumption when reporting uncertainties; to assume a normal distribution and report plus-minus two standard deviations around the mean. We would like the train/calibrate/test split to contain data from different labels, yet we do not want to sacrifice too much training data for calibration or testing. The ensemble size is seen to be sufficient for the applications presented here, but for more realistic potentials trained on more diverse datasets and more challenging QoIs, a rigorous test of convergence would be necessary to establish a minimum number of samples to balance between statistical need and computational effort, as well as identfiying any QoI difficulty dependence - i.e. for more difficult QoIs, are more ensemble members required to obtain a useful heuristic uncertainty and accurate `mean' value.

\section{Results}

To review the workflow, for a given QoI, and model $V$ with dataset $(\mathbf{x},\mathbf{y})$, we train an ACE potential on a subset of the data $(\mathbf{x}_{\text{train}},\mathbf{y}_{\text{train}})$, optimise the precision hyperparameters $\alpha$ and $\beta$ to maximise the evidence and form an ensemble of potentials drawn from our posterior distribution for coefficients. The potentials defined by these coefficients are pushed through the QoI simulation to obtain a mean prediction $\mu$ and heuristic estimate of uncertainty $\sigma$ (in our case, the standard deviation). We then calculate the $\hat{q}$ value using our calibration set $(\mathbf{x}_{\text{cal}},\mathbf{y}_{\text{cal}})$ of energy, force and virial stress observations, and form a prediction set $\mathcal{C}$ on the predicted QoI.

We will evaluate the uncertainty in the results of  four different QoIs of increasing complexity, investigating the effect of increasing the amount of data available for training and calibration, and increasing the complexity/size of the potential basis. 

\subsection{Bulk modulus}

We begin by estimating the uncertainty in predictions of the bulk modulus $B$ of cubic silicon. The calculation of this QoI is simple as it does not require any geometry relaxations or dynamics, we simply shrink or expand the unit cell around the minimum, and perform a cubic fit the bulk modulus to the curvature of the calculated energy-volume curve. This property is relatively straightforward to achieve a result with only bulk configurations, and for this reason it is the only QoI for which we will also evaluate a matrix of different models trained on varying amounts of input DFT data. 

Utilising a dataset of 489 diamond-cubic Si configurations with DFT energy, force and virial observations, we always calibrate on the same calibration set of 39 configurations. The training set varies in size from 25 to 350 DFT configurations. The linear potentials vary in size from $p_{\text{max}} = 3 - 24$, or $9 - 4330$ parameters.

\begin{figure}[!htbp]
	\centering
	\includegraphics[width=0.5\textwidth]{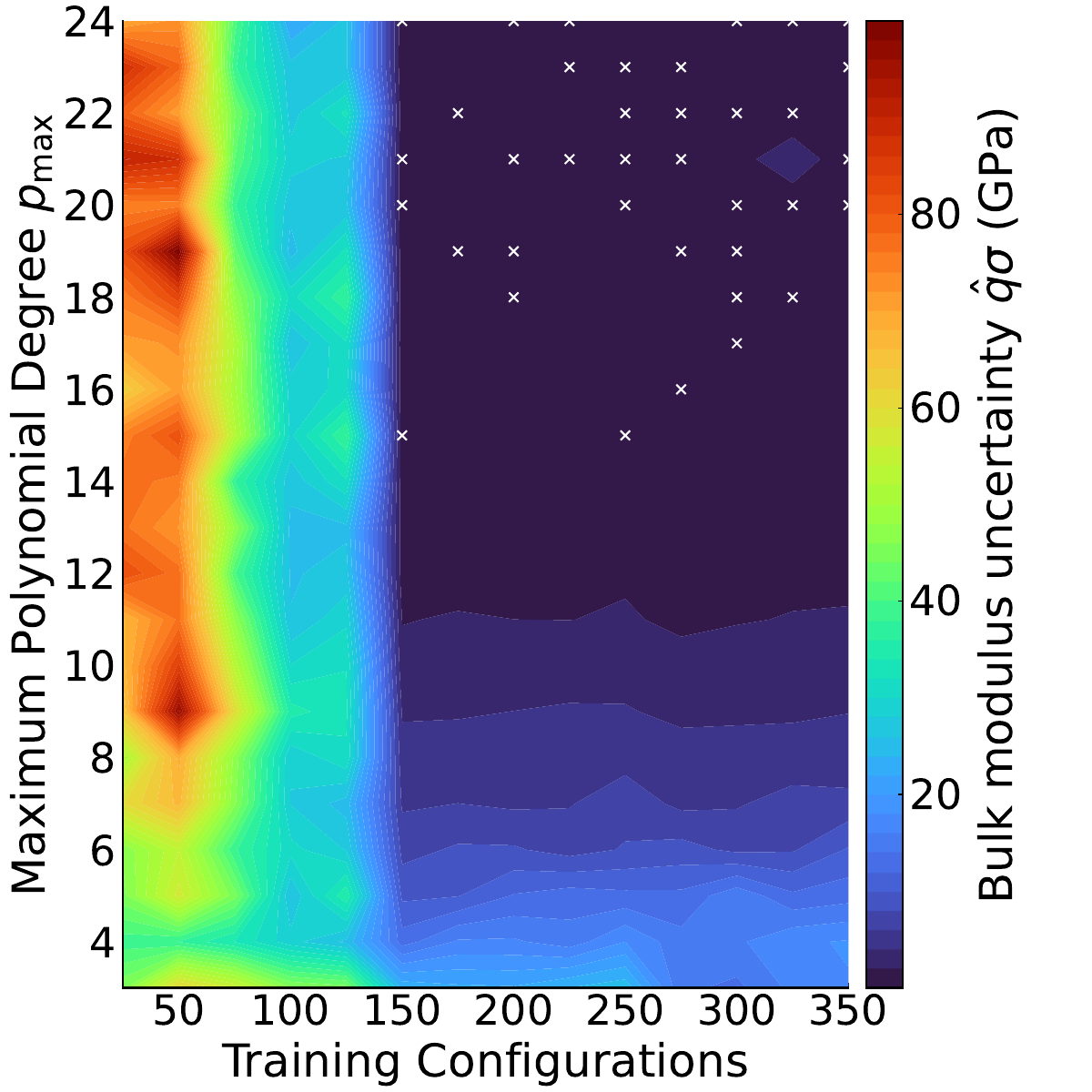}
	\caption{Conformal uncertainty $\hat{q}\sigma$ bound for predictions of bulk modulus of Si, for ACE potentials with $p_{\text{max}} = 3 - 24$ (between 9 and 4330 parameters) and trained on between 25 and 350 DFT structures of bulk Si, with those structures ranging in size from 2 to 128 atoms. White crosses indicate points where the DFT reference value lies outside the predicted range.}
	\label{Bulk modulus conformal uncertainty}
\end{figure}

An overview contour plot of the (data, complexity, uncertainty) space is presented in Figure \ref{Bulk modulus conformal uncertainty}, where we show the half bound of size $\hat{q}\sigma_{\text{pred}}$ across the entire space - see Table \ref{bulk modulus selected values} for selected predictions and uncertainties in $B$. As expected, in the low data, simple potential regime, the uncertainty in our predictions of $B$ are large. Training potentials on more data (slicing Figure \ref{Bulk modulus conformal uncertainty} from a point on the $y-$axis) generally improves the prediction and decreases the uncertainty in $B$, but with diminishing improvements past $\sim$150 atomic configurations. Similarly for a large enough training set, increasing the size of the potential (by slicing upwards from a given number of configurations on the $x-$axis) also improves the prediction and decreases the uncertainty in $B$. However for small training datasets we see some overfit, leading to uncertainties that can grow rapidly, and for large datasets the predictions can become overconfident, as represented by the white crosses in Figure \ref{Bulk modulus conformal uncertainty}.

The reference DFT value, which was not included in the training or calibration sets, is contained within the uncertainty bounds of a majority of the space of Figure \ref{Bulk modulus conformal uncertainty}, and for those for which the predictions do not agree, the discrepancy is usually a small amount. This conservative behaviour is a desirable feature: if we have constructed our model such that the variation in ensemble members captures the behaviour of the underlying uncertainty in a QoI, and conformalized the uncertainty correctly, the prediction bounds should contain the true values of the energies, forces and virials of the new structures used in the QoI calculation with probability close to $1-\zeta$ - and therefore we expect that the uncertainty in our QoI is also a good estimate of the true error.

\begin{figure*}[!htbp]
	\centering
	\begin{subfigure}[b]{0.49\textwidth}
		\centering
		\includegraphics[width=\textwidth]{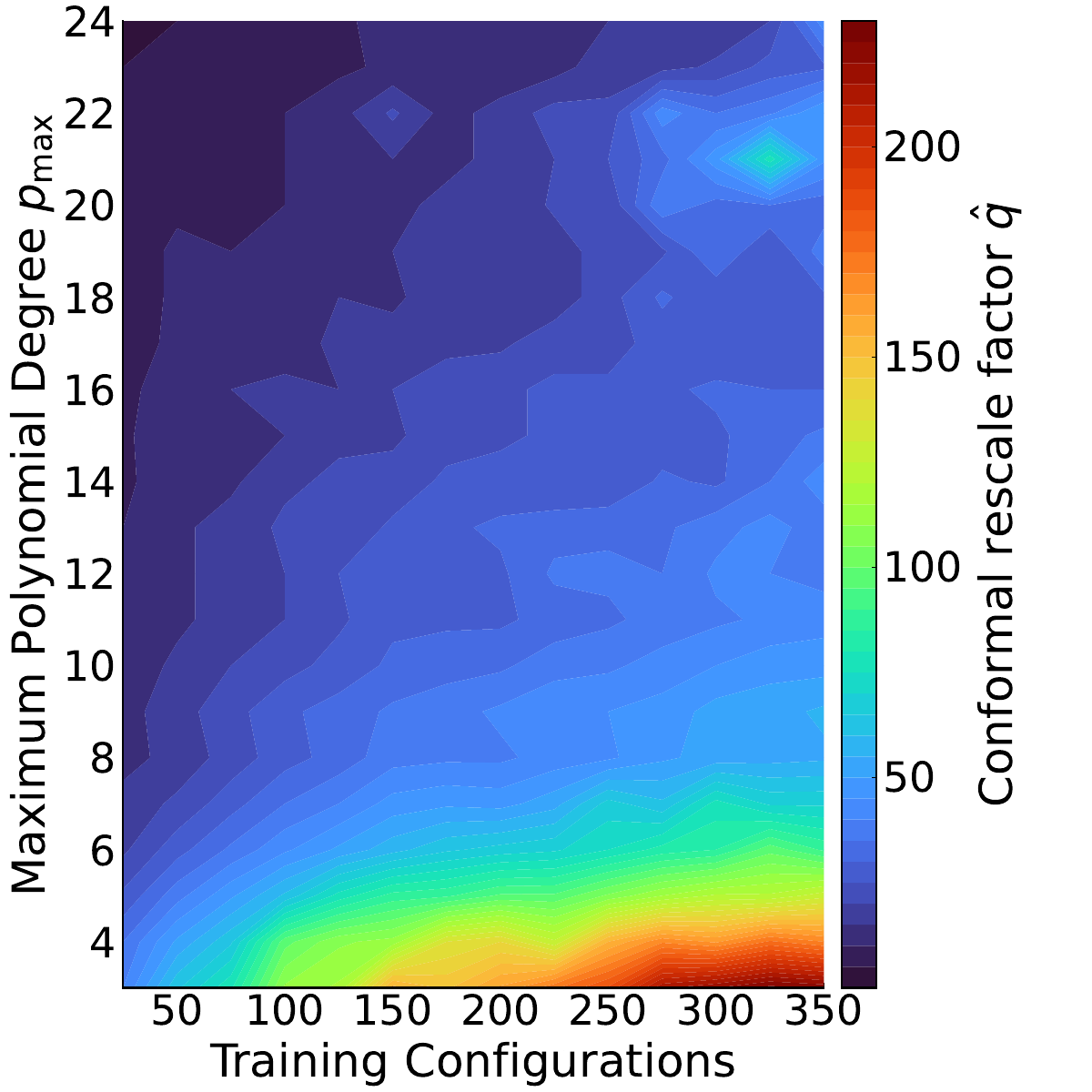}
	\end{subfigure}
	\hfill
	\begin{subfigure}[b]{0.49\textwidth}
		\centering
	\includegraphics[width=\textwidth]{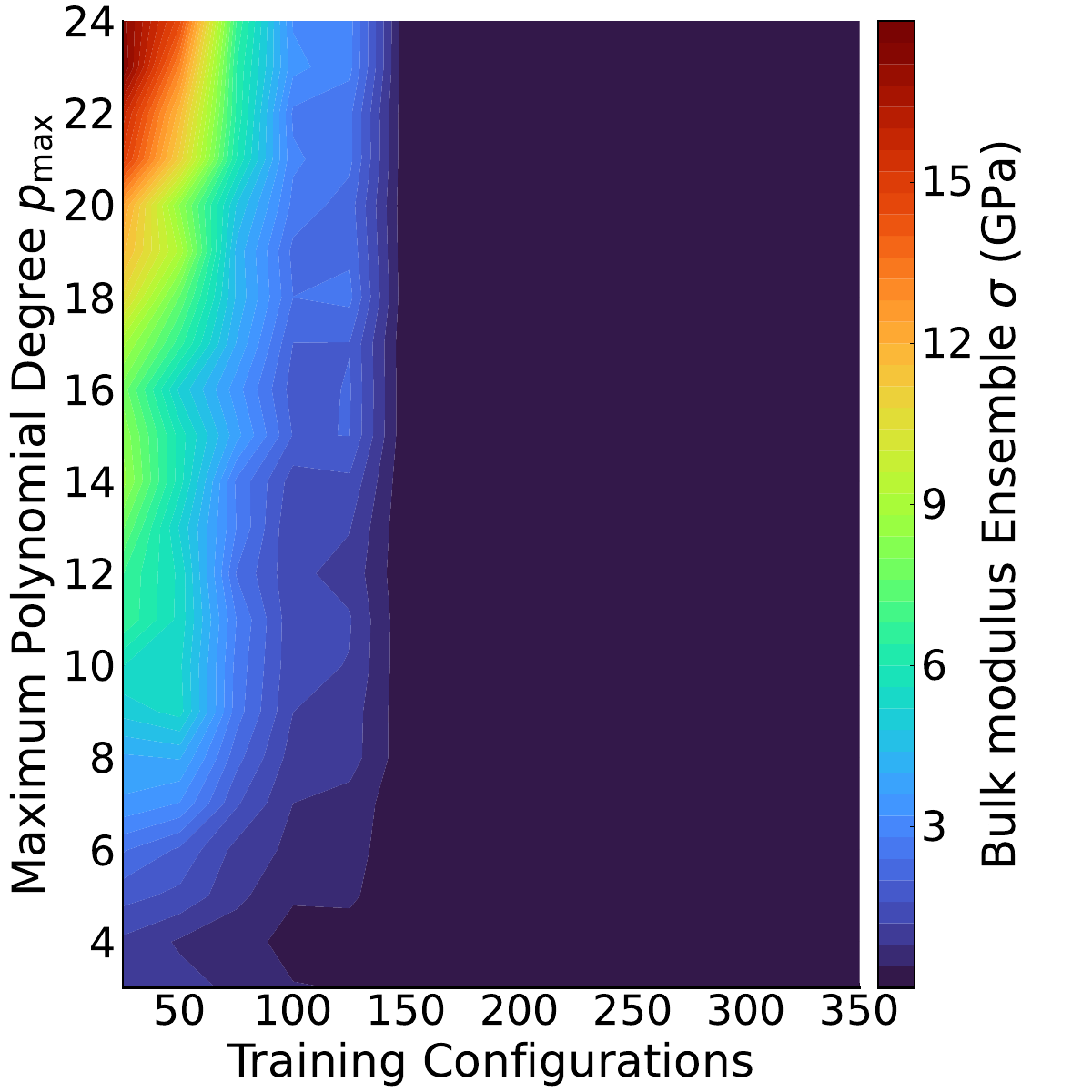}
	\end{subfigure}
	\caption{Left: conformal rescale factor $\hat{q}$ values for various potentials and training data used to calculate bulk modulus in Figure \ref{Bulk modulus conformal uncertainty}. Right: ensemble standard deviation $\sigma$ for $B$ across the same space. The product of these two heatmaps gives Figure \ref{Bulk modulus conformal uncertainty}. Note the different colourbar scales for the two subplots.}
	\label{qhat and ensemble sigma decomposition}
\end{figure*}

We can decompose the conformal uncertainty of our predictions of $B$ from Figure \ref{Bulk modulus conformal uncertainty} into the components $\hat{q}$ and the ensemble standard deviation $\sigma$; this is presented in Figure \ref{qhat and ensemble sigma decomposition}. Regardless of the size of the basis, the value of the ensemble $\sigma$ for $B$ becomes very small once the potentials are trained on more than 150 configurations. Adding more of the same type of configuration past this point does not lead to a large change in the value of the parameters - the weight posterior (Equation \ref{weight posterior}) becomes very narrow, and the ensemble members become very similar. The largest $\sigma$ values occur for potentials with large basis sets (high maximum polynomial degree) and few training configurations, which we have seen the effect of previously in Figure \ref{Bulk modulus conformal uncertainty}.

The conformal $\hat{q}$ scale factor gives the largest corrections to simple potentials trained on large amounts of data. Since we become more certain of our hyperparameters $\alpha$ and $\beta$ as we train on more data for a fixed basis, and the posterior distribution of weights becomes narrower, so there is less variation in predictions on the atomistic quantities (energies, forces and virials) contained in the calibration set. For the simple IPs, where predictions are likely to be less good, the corresponding scores (Equation \ref{score function}) will be larger and hence $\hat{q}$ must become larger.

We can focus on the top right region of the heatmaps from Figures \ref{Bulk modulus conformal uncertainty} and \ref{qhat and ensemble sigma decomposition} - models trained on larger amounts of data and with more parameters to fit - to observe the effect of overfitting / overconfidence in our uncertainties. In this region, since our ensemble members do not vary greatly, we rely on the $\hat{q}$ corrective factor - calculated by assessing performance on the quantities the potentials are trained on - to capture the QoI uncertainty. We can pick out individual points whose predictions agree with DFT (e.g. $(\# \mathrm{Configs},p_{\text{max}}) = (150,22),(325,21)$) and attribute those to a more fair assessment of the $\hat{q}$ factor for those potentials'; nevertheless these points are outliers. Increasing the number of ensemble members, assessing the performance of the chosen score function, and training on a more diverse dataset in this region would all be suitable next steps to rectify these disagreements and improve the uncertainty estimates.

\subsection{Elastic constants}

Moving up a step in QoI complexity from calculation of bulk moduli, we now focus on calculating elastic constants $C_{11}$, $C_{12}$ and $C_{44}$ of bulk silicon. These quantities are more complex because they require the chosen potential to perform optimisations of the atomic positions and the crystal unit cell for the initial bulk configuration, and of atomic positions for strained configurations. Using five configurations, with strain ranging between $\pm 0.025$, are then used in a linear regression scheme to calculate $C_{11}$,$C_{12}$ and $C_{44}$. These quantities are related to the bulk modulus for cubic crystals, and we therefore expect that for the suitably complex potentials trained on enough data from Figure \ref{Bulk modulus conformal uncertainty}, we should be able to capture $C_{11}$ and $C_{12}$ fairly confidently, and account for $C_{44}$ within uncertainty.

Since we have already constructed the potentials, built the design matrix $\Phi$ and vector of DFT observations $\mathbf{y}$, and performed the hyperparameter optimisations for $\alpha$ and $\beta$ for a wide range of potentials trained only on bulk configurations in our bulk modulus results above, we could perform a similar analysis here - the potentials, optimised hyperparameters and $\hat{q}$ values are QoI-agnostic, which is a desirable feature for transferable potentials. However, we choose to focus on the `data' axis of Figure \ref{Bulk modulus conformal uncertainty}, fixing the size of the potentials to $p_{\text{max}} = [4,12,19]$ corresponding to $[13, 214, 1432]$ parameters, so that we can visualize the uncertainties and predictions of the three elastic constants for a varying amount of training data.

\begin{figure*}[!htbp]
    \centering
    \includegraphics[width=0.99\textwidth]{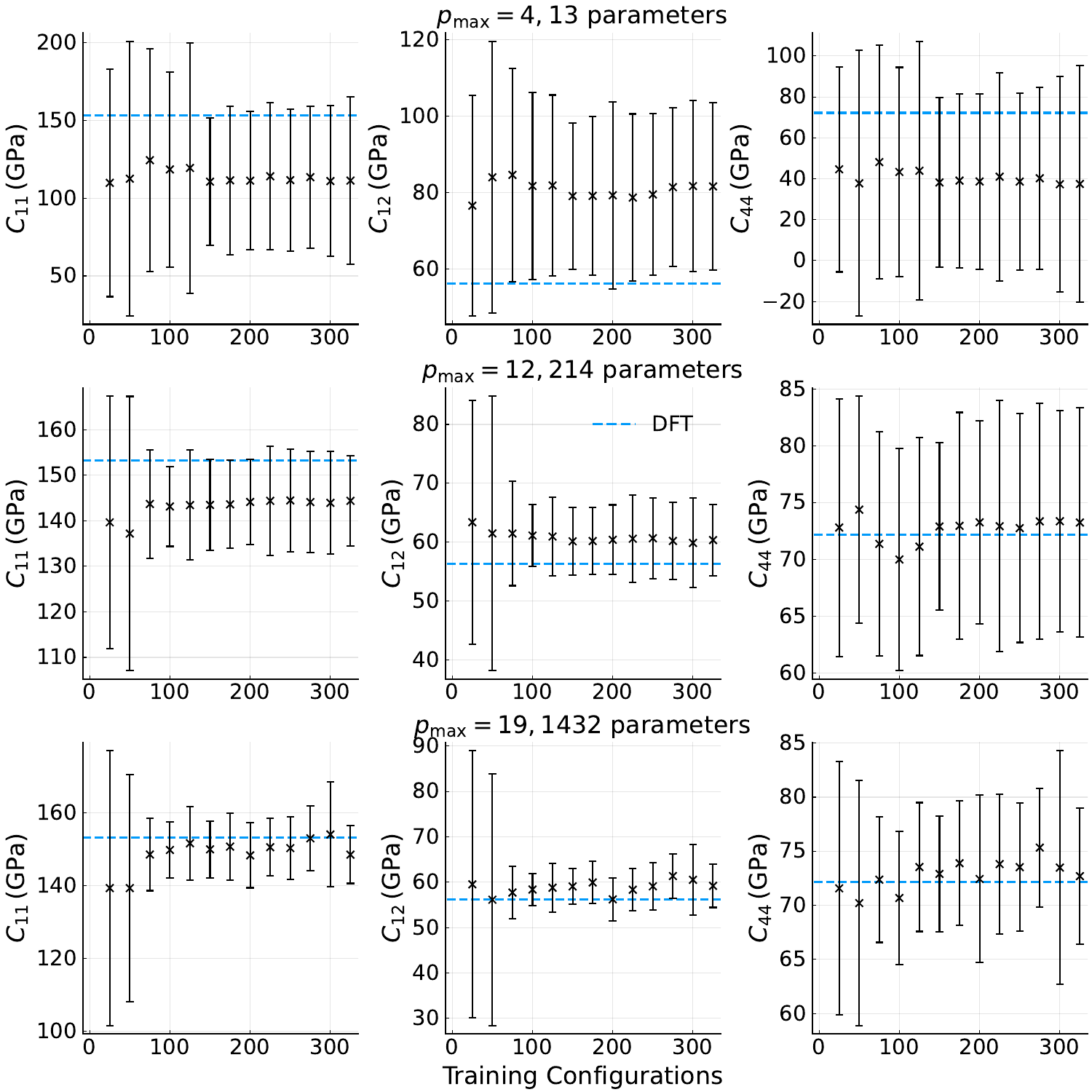}
    \caption{Mean prediction and conformal uncertainty bounds for elastic constants $C_{11}$,$C_{12}$ and $C_{44}$ of cubic Si, for three potentials with $p_{\text{max}}=[4,12,19]$ trained on varying numbers of atomic structures. The DFT reference prediction (dashed line) is captured within uncertainty across almost entire range for each potential.}
    \label{elastic constants vs configs}
\end{figure*}

As we might expect due to the relationship between the bulk modulus and elastic constants for cubic crystals, we capture $C_{11}$ and $C_{12}$ across almost the entire range of training configurations used for each potential; the same is also true for $C_{44}$ - this is shown in Figure \ref{elastic constants vs configs}, and selected values are also tabulated in Table \ref{elastic constants selected values}.

\begin{figure*}[!htbp]
    \centering
    \includegraphics[width=\textwidth]{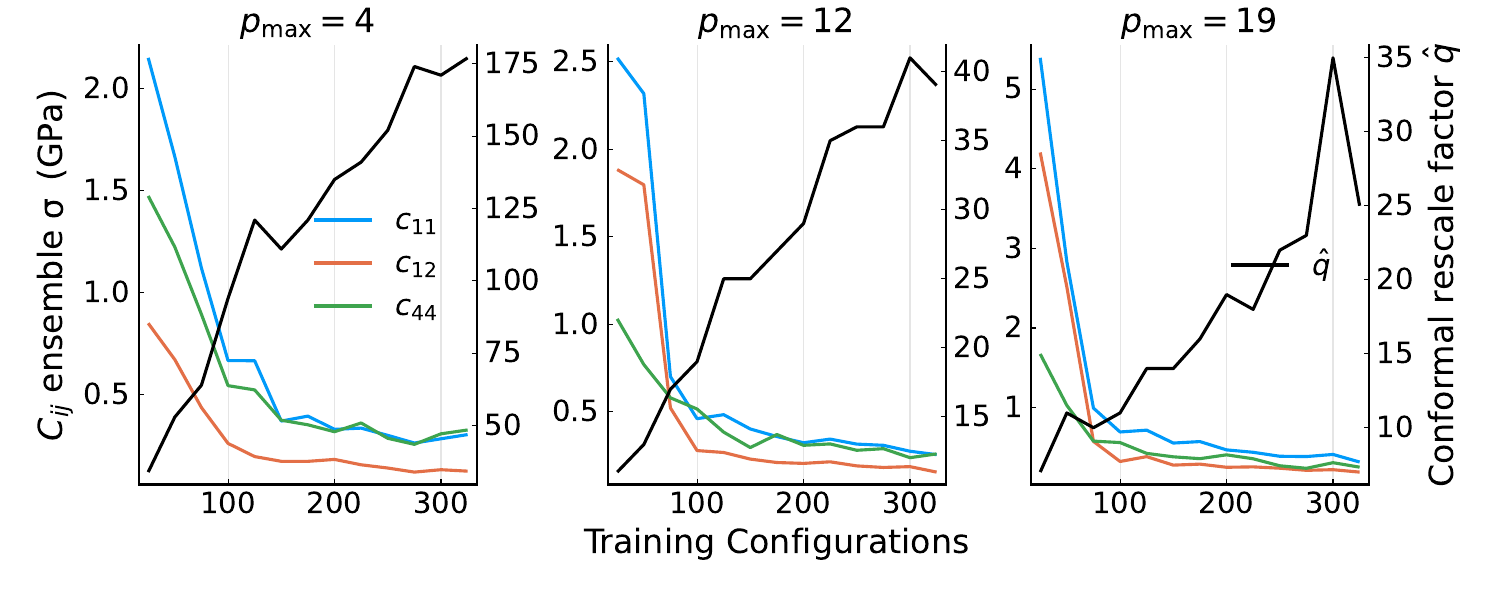}
    \caption{Ensemble standard deviations for elastic constants $C_{11}$, $C_{12}$ and $C_{44}$ on the left $y-$axes, and $\hat{q}$ factors on the right $y-$axes, for three potentials, $p_{\text{max}}=[4,12,19]$, trained on varying amounts of DFT atomic structures. Note the different scales for the three subplots. The three products of the curves from each of the subplots above correspond to half the uncertainty bounds presented in Figure \ref{elastic constants vs configs}.}
    \label{elastic constants qhat and sigmas}
\end{figure*}

The predictions for the elastic constants seem to have two distinct regions of behaviour: a region of fewer training configurations where we can infer that the uncertainty is caused by lack of data, and a remaining region of increasing training configurations where we infer that the uncertainty is dominated by lack of complexity in the potential. Interestingly, comparing Figure \ref{elastic constants vs configs} to Figure \ref{Bulk modulus conformal uncertainty}, the position of this transition changes depending on the basis size: for $p_\mathrm{max} = 4$, the top row of Figure \ref{elastic constants vs configs}, the transition occurs at approximately 150 training configurations. For the more complex potentials, the transition occurs sooner at approximately 75 training configurations. 
To investigate this further we decompose the uncertainty into $\hat{q}$ and ensemble standard deviations for $C_{11}$, $C_{12}$ and $C_{44}$, shown in Figure \ref{elastic constants qhat and sigmas}. We can identify that in the regions that are training data limited, the ensemble $\sigma$ for all the $C_{ij}$ decrease quickly as we increase the amount of training data; this decrease appears to occur faster as the complexity of the potential increases. In the region where we are limited by the complexity of the IP, the ensemble standard deviations are decreasing slowly, with the corresponding increasing $\hat{q}$ values being driven by increasingly narrow posterior weight distributions, as the model becomes more certain of the parameter values.

Not investigated here is the effect on the uncertainty estimate from the random splits into training and calibration sets, which may alter the mean prediction, ensemble standard deviation, and the multiplicative $\hat{q}$ factor. From the 489 total bulk structure data points, the number of possible ways to choose $b$ points is given by the binomial coefficient $\prescript{489}{}{C}_{b}$ - for $b$ taking values of training configurations as in Figure \ref{elastic constants vs configs}, the number of combinations are incredibly large. An interesting place for further work would be quantifying this effect; we might expect that the fine structure we see in $\hat{q}$ and $\sigma$ in Figure \ref{elastic constants qhat and sigmas}, when averaged over different train/calibration/test splits is effectively noise, leaving the underlying trend.

\subsection{Vacancy formation energy}

To increase the QoI difficulty once again, we target a quantity which requires geometry optimisations, but also needs more varied training data. The calculation of a vacancy formation energy $E_{\text{vac}}$ fits this description since the IP must accurately represent both the bulk crystal and the monovacancy configurations. 

This requires the generation of new potentials and calculation of hyperparameters (see Methodology), since we now train and calibrate on the bulk configurations used previously as well as monovacancy and divacancy configurations. In a similar, perpendicular vein to the previous section, we restrict our analysis this time to a fixed amount of training data whilst allowing the number of parameters in the potential to vary. We train on a fixed random selection of 500 structures and calibrate on a separate 62, with 156 structures held back as a test set. As outlined in the discussion on this application of conformal predictions, with this new train/calibrate/test split we do not explicitly guarantee that we have vacancy structures in the training and calibration sets, but this is not expected to be a problem for this size of subsets of data.

\begin{figure}[!htbp]
    \centering
    \includegraphics[width=0.48\textwidth]{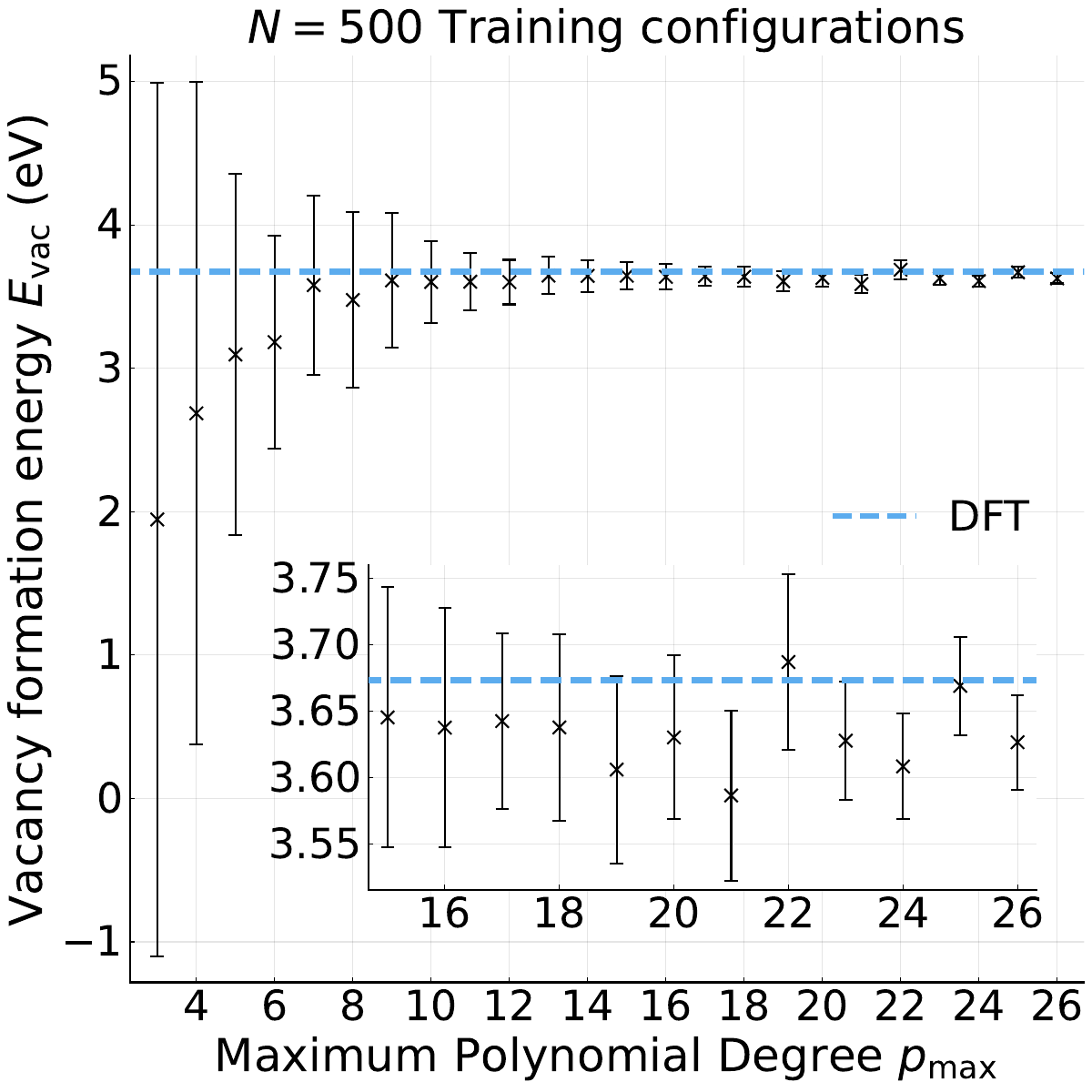}
    \caption{Mean prediction and conformal uncertainty bounds for relaxed vacancy formation energy of Si, for potentials of increasing size trained on the same dataset of 500 bulk, mono- and di-vacancy DFT structures, and calibrated on 62 similar structures. The dashed line is the DFT reference result \cite{Bartok2018}, $E_{\text{vac}}^{\text{DFT}} = 3.67 \,\,\text{eV}$.}
    \label{vfe vs complexity}
\end{figure}

Results for the vacancy formation energy are shown in Figure \ref{vfe vs complexity}. The simpler potentials with fewer available parameters to be trained on the data give predictions that are both far from the reference result, and with large uncertainty bounds. Note that prediction sets which include negative values for $E_{\text{vac}}$ do not truly observe negative values - instead the conformal scaling is large since the model is very simple and does not perform well (i.e. training, calibration and test errors will be large).

As $p_{\text{max}}$ is allowed to increase, the mean prediction of the ensemble generally comes closer to the reference value, and the uncertainty bounds shrink to a minimum value of $\pm 0.036 \,\,\text{eV}$ at $p_{\text{max}} = 26$. The conformal uncertainty routine gives coverage for the DFT result across (almost) the entire range of potentials - only for the more complex IPs with many parameters do we lose this coverage as we overfit to the training data.

\begin{figure}[!htbp]
    \centering
    \includegraphics[width=0.48\textwidth]{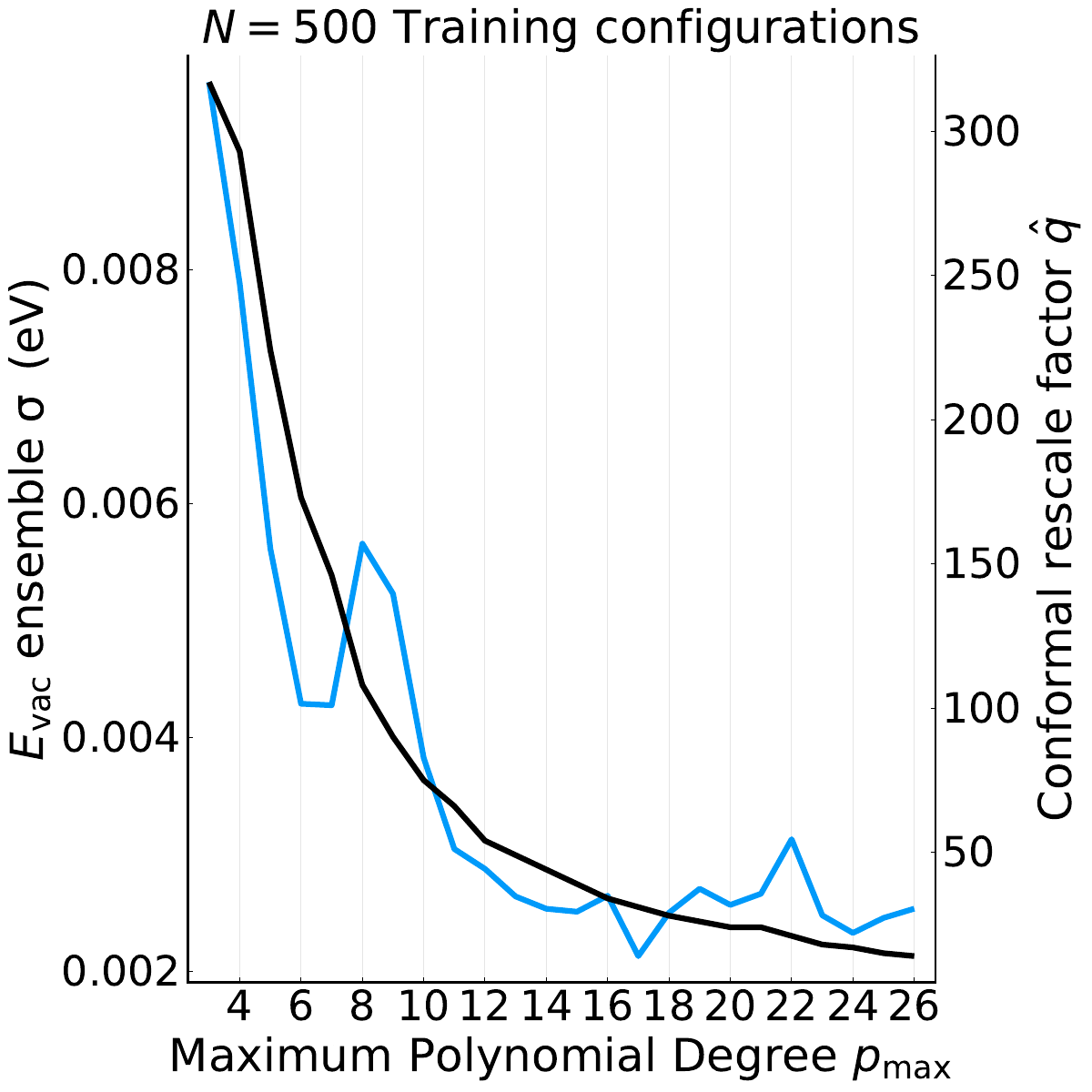}
    \caption{Ensemble standard deviations (left axis) for vacancy formation energy and $\hat{q}$ factor (right axis) for potentials ranging from $p_{\text{max}}=3-26$, (number of parameters 9--6,435) trained on 500 bulk, mono- and di-vacancy DFT structures, and calibrated on 62 similar structures. The product of the two curves produces half the uncertainty bound presented in Figure \ref{vfe vs complexity}.}
    \label{vfes qhat and sigmas}
\end{figure}

To investigate the structure of these uncertainty bounds, we again decompose the uncertainty into the $\hat{q}$ factor (evaluated on atomic properties of a calibration set), and the ensemble heuristic uncertainty for the spread of $E_{\text{vac}}$ predictions; this is presented in Figure \ref{vfes qhat and sigmas}. We observe that both $\hat{q}$ and the ensemble standard deviation $\sigma$ initially decrease as the model gains more parameters, since the increased flexibility allows for better representation of the calibration set and for more sensible relaxations and energy/force evaluations to be done on both the bulk and monovacancy structures during evaluation of $E_{\text{vac}}$. This decrease slows as we continue increasing the number of parameters, and the $\sigma$ value stagnates to a roughly constant value at approximately $p_{\text{max}} = 15$, suggesting that we have crossed from a region where we are limited by our model to a region where we are limited by the available training data. For $\hat{q}$ we see the rate of decrease slow as $p_{\text{max}}$ increases, suggesting that the performance on the calibration set (see Equations \ref{score function}-\ref{q hat and q val}) is continuing to improve as the basis becomes larger, even if this improvement is not represented in $\sigma$. 

Even though the performance on our calibration set suggests that we could increase the size of our IP further without risk of overfitting, the behaviour of the heuristic ensemble $\sigma$ implies that we require more training data. An interesting possible idea for determining whether to gather more training data or increase the size of a single IP would be to vary the amount of training data and size of the given potential, mapping the heuristic uncertainties $\sigma$ for some QoI(s) and the calculated $\hat{q}$ values, and then inspecting similar graphs to Figures \ref{elastic constants qhat and sigmas} and \ref{vfes qhat and sigmas}.

\subsection{Vacancy migration}

We target a vacancy migration in silicon as our final QoI. This property is once again a step up in terms of difficulty, as it requires the calculation of an energy barrier $E_{\text{b}}$ and the energies along the path as the configuration moves from the initial to the final state. This is done using the Nudged Elastic Band (NEB) technique - we set up a series of images between the initial and final states, and by minimising the energy between them whilst maintaining the spacing between the images via spring forces, the minimum energy path is produced. This is performed using the ASE \cite{ase-paper} NEB class, with 11 total images inclusive of the initial and final vacancies.

As QoIs become more complex, the chosen potential is more likely to be extrapolating outside of environments it has been trained and calibrated on. This means that we inherently have less intuition \textit{a priori} for which types of data we need to include during training to obtain sensible predictions in target QoIs. This logic also applies to the calculation of uncertainties: if we calibrate uncertainties on some configurations, but during the process of a QoI calculation we have needed to extrapolate to unseen regions in the PES, we cannot then be certain that the calibration of uncertainties will be meaningful. This highlights the need for either carefully curated datasets, or some automated active learning scheme based on UQ: both prediction and uncertainty are dependent on the training and calibration sets being (at least close to) representative of the new points on which we predict in the course of a simulation.

\begin{figure}[!htbp]
    \centering
    \includegraphics[width=0.48\textwidth]{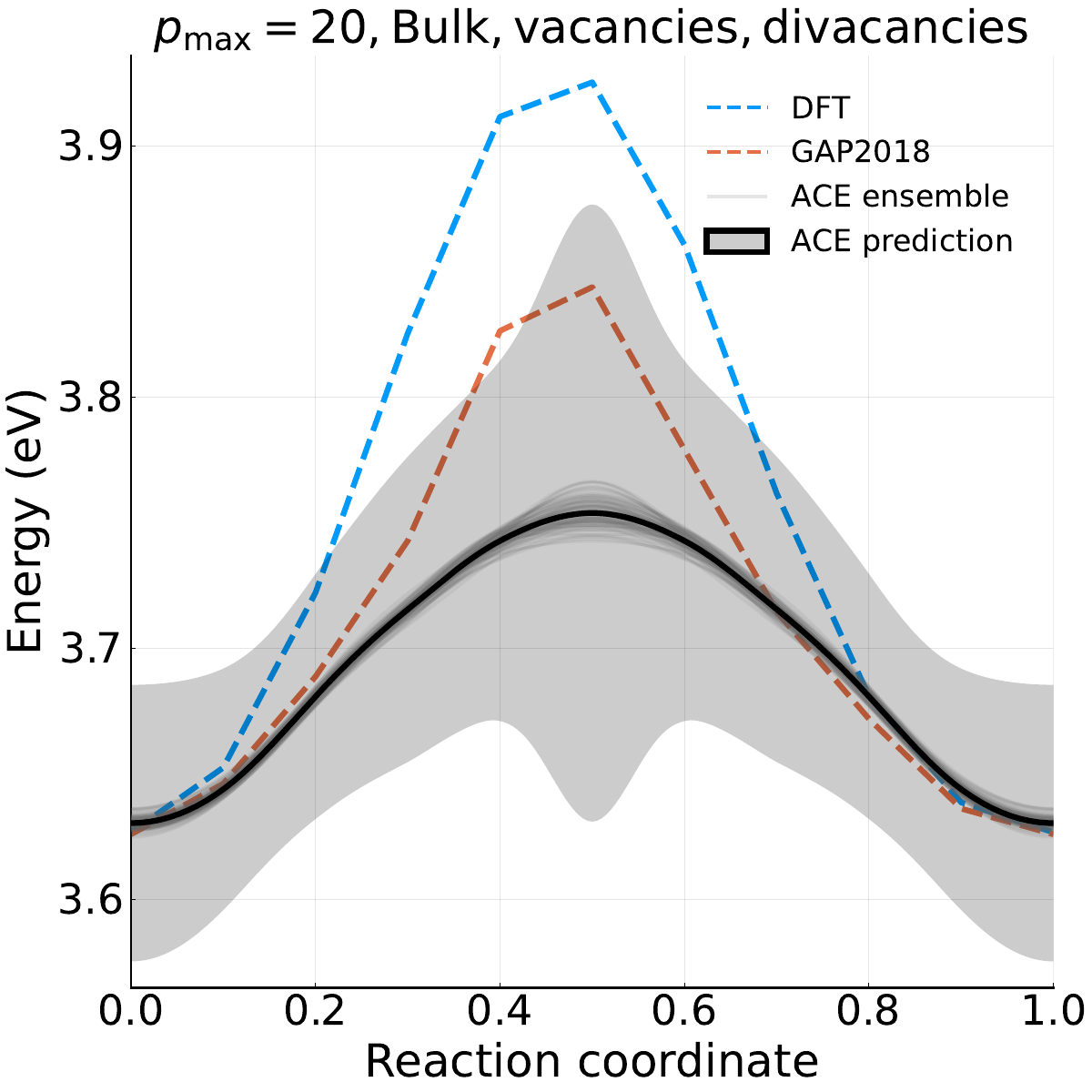}
    \caption{Single vacancy migration in silicon for an ACE potential with $p_{\text{max}}=20$ (1821 parameters) trained on 500 bulk, mono- and di-vacancy DFT structures, showing samples drawn from the weight posterior (grey curves), the mean prediction (black curve) and conformalized uncertainty (grey ribbon), with the DFT reference curve (dashed blue) and the Si GAP prediction \cite{Bartok2018} (trained on the full dataset) for comparison.}
    \label{bulk vac divac migration}
\end{figure}

Using one of the potentials from the previous section ($p_{\text{max}} = 20$, trained on the same data) we obtain an ensemble of minimum energy paths for a single vacancy migration in Figure \ref{bulk vac divac migration}, showing the central prediction as well as the conformal uncertainty bound. Since the calibration set is fixed (and hence so is the value for $\hat{q}$), the changes in the uncertainty across the path are driven by the spread of the ensemble members. At the initial and final points - which correspond to mono-vacancy configurations - we are very close to the DFT result and the uncertainty is small, which is expected since the bulk structure and vacancies are well represented in the dataset. The model is most uncertain at the saddle point, which matches our intuition since our training and calibration sets do not include these configurations: the potential is being forced to extrapolate. We can see that all members of the ensemble find an energy barrier, and whilst the uncertainty bound show a possible minimum at the saddle point, this is not truly observed and is a symptom of the large spread of predictions there - this is similar to poor models in Figure \ref{vfe vs complexity} having uncertainty bounds which `predict' negative vacancy formation energies.

The DFT result gives a barrier height $E_{\text{b}}^{\text{DFT}} = 0.30 \,\,\text{eV}$ and a vacancy formation energy (which correspond to the initial and final images of the paths) of $E_{\text{vac}}^{\text{DFT}} = 3.63 \,\,\text{eV}$, whereas the ACE model predicts $E_{\text{b}} = 0.12 \pm 0.12\,\,\text{eV}$ and $E_{\text{vac}} = 3.63 \pm 0.06\,\,\text{eV}$. For such a relatively simple potential, trained and calibrated on a cut-down portion of the full DFT dataset, both the prediction of the energy barrier and the fact that our uncertainty bound is close to the DFT across the predicted, symmetric minimum energy path is quite impressive. The full Si GAP potential \cite{Bartok2018} also struggles to reproduce the DFT barrier here, predicting a barrier height $E_{\text{b}}^{\text{GAP}} = 0.22$ eV.

\section{Conclusion}

We have explored a new framework for quantifying the uncertainties in different QoIs for pure silicon, for a range of different size of ACE IPs and amounts of training data. This new methodology combines a base, Bayesian forward propagation method incorporating parametric uncertainty and uncertainty in the training data, with a frequentist conformal prediction to generate robust prediction sets on physical quantities outside of the training and calibration set. By comparison to reference DFT results for a range of QoIs, we observe good coverage for many of the constructed potentials. We further investigate the structure of these error bars by decomposition into the constituent quantities; the ensemble standard deviation and the conformal rescale factor.

\section{Software Availability}

Code and data supporting this work will be made available on publication.

\begin{acknowledgments}
I.B. is supported by a studentship
within the UK Engineering and Physical Sciences Research Council-supported Centre (EPSRC) for Doctoral Train-
ing in Modelling of Heterogeneous Systems, Grant
No. EP/S022848/1. J.R.K. acknowledges funding
from the Leverhulme Trust under grant RPG-2017-191 and the NOMAD Centre of Excellence funded by the European Commission under grant agreement 951786.
Computing facilities were provided by the Scientific Computing Research Technology Platform at the University of Warwick.

The authors acknowledge insightful discussions with Ryan S. Elliot  and Mark Transtrum in the context of the partnership between the HetSys CDT and OpenKIM and with Ralf Drautz and Yury Lysogorskiy in the context of the partnership between HetSys and ICAMS.
\end{acknowledgments}

\section{References}
\bibliography{references}
\newpage
\appendix*
\section{}


\begin{table*}[h!]
\begin{center}
    \begin{tabular}{|c|c|c|c|c|} 
 \hline
 $p_{\text{max}}$ $\backslash$ \# Configs & 50 & 150 & 250 & 350 \\ [0.5ex] 
 \hline
 4 & 93.42 ± 37.9 & 89.54 ± 13.08 & 90.24 ± 18.01 & 91.06 ± 18.18 \\ 
 \hline
 9 & 85.44 ± 96.6 & 88.01 ± 3.81 & 87.44 ± 4.13 & 87.66 ± 3.97 \\ 
 \hline
 13 & 85.02 ± 71.84 & 88.58 ± 1.11 & 88.26 ± 1.47 & 88.47 ± 1.28 \\ 
 \hline
 19 & 75.75 ± 99.89 & 88.77 ± 0.65 & 89.09 ± 0.95 & 88.33 ± 1.05 \\ 
 \hline
 24 & 59.27 ± 72.75 & \textbf{89.44 ± 0.47} & 88.75 ± 0.57 & \textbf{91.3 ± 1.02} \\
 \hline
\end{tabular}
\end{center}
\caption{Selected predictions for Si bulk modulus with uncertainties from Figure \ref{Bulk modulus conformal uncertainty}. All values are reported in units of GPa, and bold values are those which do not agree with the DFT reference result \cite{Bartok2018} of $B_{\text{DFT}} = 88.6$ GPa.}
\label{bulk modulus selected values}
\end{table*}

\begin{table*}[h!]
\begin{center}
    \begin{tabular}{|c|c|c|c|c|} 
 \hline
 $p_{\text{max}}$ $\backslash$ \# Configs & 25 & 125 & 225 & 325 \\ [0.5ex] 
 \hline
 4 & \begin{tabular}{c} 109.85 ± 73.18 \\ 76.6 ± 28.89 \\ 44.61 ± 50.13 \end{tabular} & \begin{tabular}{c} 119.42 ± 80.57 \\ \textbf{81.88 ± 23.65} \\ 43.89 ± 63.24 \end{tabular} & \begin{tabular}{c} 114.07 ± 47.25 \\ \textbf{78.71 ± 21.87} \\ 40.98 ± 50.81 \end{tabular} & \begin{tabular}{c} 111.29 ± 53.71 \\ \textbf{81.62 ± 21.93} \\ 37.45 ± 57.71 \end{tabular} \\ 
 \hline
 12 & \begin{tabular}{c} 139.66 ± 27.75 \\ 63.37 ± 20.72 \\ 72.81 ± 11.33 \end{tabular} & \begin{tabular}{c} 143.44 ± 12.07 \\ 60.94 ± 6.67 \\ 71.13 ± 9.58 \end{tabular} & \begin{tabular}{c} 144.39 ± 12.02 \\ 60.57 ± 7.48 \\ 72.92 ± 11.06 \end{tabular} & \begin{tabular}{c} 144.39 ± 9.92 \\ 60.34 ± 6.04 \\ 73.25 ± 10.1 \end{tabular} \\ 
 \hline
 19 & \begin{tabular}{c} 139.34 ± 37.75 \\ 59.53 ± 29.43 \\ 71.57 ± 11.72 \end{tabular} & \begin{tabular}{c} 151.65 ± 10.04 \\ 58.79 ± 5.38 \\ 73.54 ± 5.95 \end{tabular} & \begin{tabular}{c} 150.57 ± 7.9 \\ 58.34 ± 4.61 \\ 73.82 ± 6.44 \end{tabular} & \begin{tabular}{c} 148.53 ± 7.91 \\ 59.19 ± 4.76 \\ 72.7 ± 6.31 \end{tabular} \\ 
 \hline
\end{tabular}
\end{center}
\caption{Selected predictions for Si $C_{ij}$ with uncertainties from Figure \ref{elastic constants vs configs}, with values listed in the order $C_{11}$, $C_{12}$, $C_{44}$ in each cell. All values are reported in units of GPa, and bold values are those which do not agree with the DFT reference results \cite{Bartok2018}, $C_{11} = 153.3$ GPa, $C_{12} = 56.3$ GPa, $C_{44} = 72.2$ GPa.}
\label{elastic constants selected values}
\end{table*}

\newpage

\end{document}